\documentclass[11pt,fleqn]{article}

\usepackage{a4}
\usepackage{amstext}
\usepackage{amsfonts}
\usepackage{amssymb}
\usepackage{color}
\usepackage{epsfig}

\newcommand{\ltapprox}{\raisebox{-0.5ex}{$\,\stackrel{<}{\scriptstyle\sim}\,$}}

\begin{document}


\begin{center}

{\huge \bf The continuum limit of the static-light}

{\huge \bf meson spectrum}

\vspace{0.5cm}

SFB/CPP-10-30, LTH 869, IFT-UAM/CSIC-10-14, HU-EP-10/20

\vspace{0.5cm}

\textbf{Chris Michael} \\
Theoretical Physics Division, Department of Mathematical Sciences, University of Liverpool, Liverpool L69 3BX, UK \\

\vspace{0.3cm}

\textbf{Andrea Shindler} \\
Istituto de F\'{\i}sica Te\'orica UAM/CSIC, Universidad Aut\'onoma de Madrid, Cantoblanco E-28049 Madrid, Spain

\vspace{0.3cm}

\textbf{Marc Wagner} \\
Humboldt-Universit\"at zu Berlin, Institut f\"ur Physik, Newtonstra{\ss}e 15, D-12489 Berlin, Germany  \\

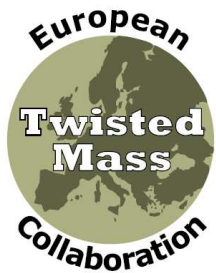
\vspace{0.5cm}

\begin{picture}(0,0)%
\includegraphics{Logo.pstex}%
\end{picture}%
\setlength{\unitlength}{4144sp}%
\begingroup\makeatletter\ifx\SetFigFont\undefined%
\gdef\SetFigFont#1#2#3#4#5{%
  \reset@font\fontsize{#1}{#2pt}%
  \fontfamily{#3}\fontseries{#4}\fontshape{#5}%
  \selectfont}%
\fi\endgroup%
\begin{picture}(1620,1620)(1,-781)
\end{picture}%

\vspace{0.4cm}

April~23, 2010

\end{center}

\vspace{0.1cm}

\begin{tabular*}{16cm}{l@{\extracolsep{\fill}}r} \hline \end{tabular*}

\vspace{-0.4cm}
\begin{center} \textbf{Abstract} \end{center}
\vspace{-0.4cm}

 We investigate the continuum limit of the low lying static-light meson
spectrum using Wilson twisted mass lattice QCD with $N_f = 2$ dynamical
quark flavours. We consider three values of the lattice spacing $a
\approx 0.051 \, \textrm{fm} \, , \, 0.064 \, \textrm{fm} \, , \, 0.080
\, \textrm{fm}$ and various values of the pion mass in the range $280 \,
\textrm{MeV} \ltapprox m_\textrm{PS} \ltapprox 640 \, \textrm{MeV}$. We
present results in the continuum limit for light cloud angular momentum
$j = 1/2 \, , \, 3/2 \, , \, 5/2$ and for parity $\mathcal{P} = + \,
, \, -$. We extrapolate our results to physical quark masses, make
predictions regarding the spectrum of $B$ and $B_s$ mesons and compare
with available experimental results.

\begin{tabular*}{16cm}{l@{\extracolsep{\fill}}r} \hline \end{tabular*}

\thispagestyle{empty}


\newpage

\setcounter{page}{1}

\section{\label{SEC1}Introduction}

A systematic way to study $B$ and $B_s$ mesons from first principles is
with lattice QCD. Since $a m_b > 1$ at currently available lattice
spacings for large volume simulations,  one needs to use for the $b$
quark a formalism such as Heavy Quark Effective Theory (HQET)
\cite{Neubert:1993mb,Mannel:1997ky} or Non-Relativistic QCD
\cite{Thacker:1990bm}. An alternative procedure has recently been 
proposed~\cite{Blossier:2009hg} which is based on HQET but does not make use of the static point.
Here we follow the standard HQET route,  which enables all sources of systematic error to be controlled. 

In the static limit a heavy-light meson will be the ``hydrogen atom'' of
QCD.  Since in this limit there are no interactions involving the heavy
quark spin, states are doubly degenerate, i.e. there is no hyperfine
splitting.
 Therefore, it is common to label static-light mesons by parity
$\mathcal{P}$ and the total angular momentum of the light degrees of
freedom $j$ with $j= |l \pm 1/2|$, where $l$ and $\pm 1/2$ denote
respectively angular momentum and spin.
 An equivalent notation is given by $l_{\pm}$, which reads  $S \equiv
(1/2)^-$, $P_- \equiv (1/2)^+$, $P_+ \equiv (3/2)^+$, $D_- \equiv
(3/2)^-$, $D_+ \equiv (5/2)^-$, $F_- \equiv (5/2)^+$, $F_+ \equiv
(7/2)^+$ ... The total angular momentum of a static-light meson is
either $J = j + 1/2$ or $J = j - 1/2$, where both states are of the same
mass. Note that in contrast to parity, charge conjugation is not a good
quantum number,  since static-light mesons are made from non-identical
quarks.

The static-light meson spectrum has been studied comprehensively by
lattice methods in the quenched  approximation with a rather coarse
lattice spacing \cite{Michael:1998sg}. Lattice studies with $N_f=2$
flavours of dynamical sea quarks have also explored this spectrum
\cite{Green:2003zza,Burch:2006mb,Koponen:2007fe,Foley:2007ui,Koponen:2007nr,Burch:2007xy,Burch:2008qx}. Here following our initial study \cite{Jansen:2008ht,Jansen:2008si}, we use $N_f=2$ flavours
and are able to reach lighter dynamical quark masses, which are closer
to the physical $u/d$  quark mass, so enabling a more reliable
extrapolation. 
 Note that in our formalism, maximally twisted mass lattice QCD,  mass
differences in the static-light spectrum are $\mathcal{O}(a)$ improved so that
the continuum limit is more readily accessible.
 We now extend our study to include three different lattice spacings, which 
gives us confidence that we are indeed extracting the continuum limit. 

 In this paper, we approach the $B$ meson spectrum by concentrating on
the unitary sector, where valence quarks and sea quarks are of the same
mass.  This is appropriate for static-light mesons with a light quark,
which is $u$ or $d$. 

 We also estimate masses of $B_s$ mesons with $s$ quarks of physical
mass, where the $s$ quark is treated as a valence quark in the sea of
light $u$ and $d$ quarks (so this is a  partially quenched study). We
took our $s$ quark mass values from ETMC studies of strange mesons
\cite{Blossier:2007vv,Blossier:2009bx}.

 Within the twisted mass formalism, it is feasible to use $N_f = 2+1+1$
flavours of dynamical sea quarks, which would give a more appropriate
focus on the static-strange meson spectrum if strange quark sea effects 
were significant. This is under study by ETMC.

In HQET the leading order is just the static limit. The next correction
will be of order $1/m_Q$, where $m_Q$ is the mass of the heavy quark.
This correction is expected be relatively small for $b$ quarks, but
larger for $c$ quarks. Lattice methods to evaluate these $1/m_Q$
contributions to the $B$ meson hyperfine splittings have been
established and tested in quenched studies
\cite{Bochicchio:1991cy,Guazzini:2007bu,Blossier:2009mg,Blossier:2010jk,Blossier:2010vz}.
We intend to explore these contributions using lattice techniques
subsequently. An alternative way to predict the spectrum for $B$ and
$B_s$ mesons is to interpolate between $D$ and $D_s$ states, where the
experimental spectrum is rather well known, and the static limit
obtained by lattice QCD assuming a dependence as $1/m_Q$. Thus the
splittings among $B$ and $B_s$ mesons should be approximately $m_c / m_b
\approx 1/3$ of those among the corresponding $D$ and $D_s$ mesons.

For excited $D_s$ mesons, experiment has shown that some of the states
have very narrow decay widths \cite{PDG}. This comes about, since the
hadronic transitions to $D K$ and $D_s M$ (where $M$ is a flavour
singlet mesonic system, e.g.\ $\eta'$, $\pi \pi$ or $f_0$) are not
allowed energetically. The isospin violating decay to $D_s \pi$ together
with electromagnetic decay to $D_s \gamma$  are then responsible for the
narrow width observed. A similar situation may exist for $B_s$ decays
and we investigate this here using our lattice mass determinations of
the excited states. This will enable us to predict whether narrow
excited $B_s$ mesons should be found. 

As well as exploring this issue of great interest to experiment, we
determine the excited state spectrum  of static-light mesons as fully as
possible. This will help the construction of phenomenological models and
will shed light on questions such as, whether there is an inversion of
the level ordering with $l_+$ lighter than $l_-$ at larger $l$ or for
radial excitations as has been predicted
\cite{Schnitzer:1978gq,Schnitzer:1989xr,Ebert:1997nk,Isgur:1998kr,Ebert:2009ua}.

Since we measure the spectrum for a range of values of the bare quark
mass parameter  $\mu_\mathrm{q}$ for the light quark, we could also
compare with chiral effective Lagrangians   appropriate to HQET. This
comparison would be most appropriate applied to heavy-light decay
constants in the continuum limit (see ref \cite{Blossier:2009gd}). Since
that study awaits  more precise renormalization constants, we do not
discuss it further here.

Since we have discussed the basic methods in a previous
paper \cite{Jansen:2008si}, in this paper we present only briefly the
details of our computation of static-light meson mass differences. We
give a full discussion of our extrapolation to the continuum and to
physical light quark masses. We also discuss the interpolation to
the physical $b$ quark mass.



\section{Lattice details}

We use $N_f = 2$ flavour gauge configurations produced by the European Twisted Mass Collaboration (ETMC). The gauge action is tree-level Symanzik improved \cite{Weisz:1982zw}, while the fermionic action is Wilson twisted mass at maximal twist (cf.\ e.g.\ \cite{Shindler:2007vp} and references therein). As argued in \cite{Jansen:2008si} this ensures automatic $\mathcal{O}(a)$ improvement for static-light spectral quantities, e.g.\ mass differences of static-light mesons, the quantities we are focusing on in this work.


We use three different values of the lattice spacing $a \approx 0.051 \,
\textrm{fm} \, , \, 0.064 \, \textrm{fm} \, , \, 0.080 \, \textrm{fm}$
and various values of the pion mass in the range $280 \, \textrm{MeV}
\ltapprox m_\textrm{PS} \ltapprox 640 \, \textrm{MeV}$. All lattice
volumes are big enough to fulfill $m_\textrm{PS} L > 3.2$. The ensembles
we are considering are listed in Table~\ref{TAB100}. Details regarding
the generation of gauge configurations and analysis procedures for
standard quantities (e.g.\ lattice spacing, pion mass) can be found in
\cite{Boucaud:2008xu,Baron:2009wt}.

\begin{table}[htb]
\begin{center}

\begin{tabular}{|c|c|c||c|c||c|c|}
\hline
 & & & & & & \vspace{-0.40cm} \\
        &                &                  &                      &                                   &              & \# and type \\
$\beta$ & $L^3 \times T$ & $\mu_\mathrm{q}$ & $a$ in $\textrm{fm}$ & $m_\textrm{PS}$ in $\textrm{MeV}$ & \# of gauges & of inversions \\
 & & & & & & \vspace{-0.40cm} \\
\hline
 & & & & & & \vspace{-0.40cm} \\
\hline
 & & & & & & \vspace{-0.40cm} \\
$3.90$ & $24^3 \times 48$ & $0.0040$ & $0.0801(14)$ & $336(6)$  & $1420$/$580$ & (spin)/(4rand) \\
       &                  & $0.0064$ &              & $417(7)$  & $1480$/-     & (spin)/- \\
       &                  & $0.0085$ &              & $478(8)$  & $1360$/$480$ & (spin)/(4rand) \\
       &                  & $0.0100$ &              & $517(9)$  &  $460$/$480$ & (6rand)/(4rand) \\
       &                  & $0.0150$ &              & $637(11)$ & $1000$/-     & (1rand)/- \\
 & & & & & & \vspace{-0.40cm} \\
\hline
 & & & & & & \vspace{-0.40cm} \\
$4.05$ & $32^3 \times 64$ & $0.0030$ & $0.0638(10)$ & $321(5)$ & $240$/$240$ & (4rand)/(4rand) \\
       &                  & $0.0060$ &              & $443(7)$ & $500$/$500$ & (4rand)/(4rand) \\
 & & & & & & \vspace{-0.40cm} \\
\hline
 & & & & & & \vspace{-0.40cm} \\
$4.20$ & $48^3 \times 96$ & $0.0020$ & $0.0514(8)$ & $284(5)$ & $420$/$420$ & (spin)/(4rand)\vspace{-0.40cm} \\
 & & & & & & \\
\hline
\end{tabular}

\caption{\label{TAB100}ensembles ($a$ and $m_\textrm{PS}$ have been
taken from \cite{Baron:2009wt}; \# of gauges considered for $B$/$B_s$
mesons; \# and type of inversions for $B$/$B_s$ mesons: (spin)~four spin
diluted timeslice sources on the same randomly chosen timeslice;
(1rand)~a single timeslice source on a randomly chosen timeslice;
(4rand)~four timeslice sources on four randomly chosen timeslices;
(6rand)~six timeslice sources on six randomly chosen timeslices).}

\end{center}
\end{table}

In Table~\ref{TAB100} we also list the number of gauges, on which we have
computed static-light correlation functions, and the number and type of
inversions performed to estimate light quark propagators
stochastically. Note that in contrast to our previous work
\cite{Jansen:2008ht,Jansen:2008si} we treat $B_s$ mesons in a partially
quenched approach, where the mass of the valence quark is approximately the
mass of the physical $s$ quark, taken from the study of  strange mesons
using the same configurations~\cite{Blossier:2007vv,Blossier:2009bx},
\begin{itemize}
\item $\beta = 3.90 \quad \rightarrow \quad \mu_{\mathrm{q},\textrm{valence}} = \mu_{\mathrm{q},s} = 0.022$,

\item $\beta = 4.05 \quad \rightarrow \quad \mu_{\mathrm{q},\textrm{valence}} = \mu_{\mathrm{q},s} = 0.017$,

\item $\beta = 4.20 \quad \rightarrow \quad \mu_{\mathrm{q},\textrm{valence}} = \mu_{\mathrm{q},s} = 0.015$,
\end{itemize}
while the sea is considerably lighter (cf.\ the listed $\mu_\mathrm{q}$ values in Table~\ref{TAB100}).



\section{Static-light mass differences}
\label{sec:energies}

The determination of static-light mass differences is essentially
identical to what we have done in \cite{Jansen:2008ht,Jansen:2008si}.

For each of our ensembles characterised by the gauge coupling $\beta$
and the twisted light quark mass $\mu_\mathrm{q}$ (cf.\
Table~\ref{TAB100}) and each of the lattice angular momentum
representations $A_1$, $E$ and $A_2$ we compute $6 \times 6$
static-light correlation matrices. The corresponding meson creation
operators differ in their (twisted mass) parity, in their $\gamma$
matrix structure and in their spatial size. They are precisely the same
we have been using before and are explained in detail
in \cite{Jansen:2008si}, section~3, Table~3.

From these correlation matrices we compute effective mass plateaux using
variational methods~\cite{Michael:1985ne,Blossier:2009kd} (cf.\
\cite{Blossier:2009vy} for exemplary plots showing the quality of our
plateaus). We extract mass differences by fitting constants to these
plateaus at sufficiently large temporal separations $T_\textrm{min}
\ldots T_\textrm{max}$. We determine $T_\textrm{min}$ and
$T_\textrm{max}$ by requiring that the reduced $\chi^2$ is
$\mathcal{O}(1)$. $T_\textrm{min}$ values are listed in
Table~\ref{TAB011}, while $T_\textrm{max} = 11$ for $\beta = 3.90$ and
$\beta = 4.05$ and $T_\textrm{max} = 17$ for $\beta = 4.20$ in most
cases (for some of the excited states smaller values had to be chosen,
because the signal was lost in statistical noise). Note, however, that
the choice of $T_\textrm{max}$ is essentially irrelevant for the
resulting mass (on the ``$T_\textrm{max}$ side'' of the effective mass
plateau statistical errors are rather large and, therefore, data points
only have a very weak effect on the fit). Since we are only interested
in mass differences $\Delta M(j^\mathcal{P}) = M(j^\mathcal{P}) - M(S)$,
the jackknife analysis has been applied directly to the mass difference
and not to the individual masses. The samples for $M(j^\mathcal{P})$ and
$M(S)$ entering for such a mass difference have been obtained with the
same value of $T_\textrm{min}$.

\begin{table}[htb]
\begin{center}

\begin{tabular}{|c|c||c|c|c|c|c|c||c|c|c|c|c|c|}
\hline
\multicolumn{2}{|c||}{\vspace{-0.40cm}} & \multicolumn{6}{c||}{} & \multicolumn{6}{c|}{} \\
\multicolumn{2}{|c||}{} & \multicolumn{6}{c||}{$B$ mesons} & \multicolumn{6}{c|}{$B_s$ mesons} \\
\multicolumn{2}{|c||}{\vspace{-0.40cm}} & \multicolumn{6}{c||}{} & \multicolumn{6}{c|}{} \\
\hline
 & & & & & & & & & & & & & \vspace{-0.40cm} \\
$\beta$ & $\mu_\mathrm{q}$ & $P_-$ & $P_+$ & $D_\pm$ & $D_+$ & $F_\pm$ & $S^\ast$ & $P_-$ & $P_+$ & $D_\pm$ & $D_+$ & $F_\pm$ & $S^\ast$ \\
 & & & & & & & & & & & & & \vspace{-0.40cm} \\
\hline
 & & & & & & & & & & & & & \vspace{-0.40cm} \\
\hline
 & & & & & & & & & & & & & \vspace{-0.40cm} \\
$3.90$ & $0.0040$   & $6$ & $6$ & $5$ & $4$ & $4$ & $4$   & $6$ & $6$ & $5$ & $5$ & $4$ & $4$ \\
       & $0.0064$   & $6$ & $6$ & $5$ & $4$ & $4$ & $4$   & - & - & - & - & - & - \\
       & $0.0085$   & $6$ & $6$ & $5$ & $4$ & $4$ & $4$   & $5$ & $5$ & $4$ & $4$ & $4$ & $4$ \\
       & $0.0100$   & $6$ & $6$ & $5$ & $4$ & $4$ & $4$   & $5$ & $5$ & $4$ & $4$ & $4$ & $4$ \\
       & $0.0150$   & $6$ & $6$ & $5$ & $4$ & $4$ & $4$   & - & - & - & - & - & - \\
 & & & & & & & & & & & & & \vspace{-0.40cm} \\
\hline
 & & & & & & & & & & & & & \vspace{-0.40cm} \\
$4.05$ & $0.0030$   & $7$ & $6$ & $6$ & $5$ & $5$ & $6$   & $7$ & $7$ & $6$ & $5$ & $5$ & $7$ \\
       & $0.0060$   & $7$ & $6$ & $6$ & $5$ & $5$ & $5$   & $7$ & $7$ & $6$ & $5$ & $5$ & $7$ \\
 & & & & & & & & & & & & & \vspace{-0.40cm} \\
\hline
 & & & & & & & & & & & & & \vspace{-0.40cm} \\
$4.20$ & $0.0020$   & $10$ & $10$ & $8$ & $7$ & $7$ & $9$   & $11$ & $9$ & $9$ & $8$ & $8$ & $11$\vspace{-0.40cm} \\
 & & & & & & & & & & & & & \\
\hline
\end{tabular}

\caption{\label{TAB011}$T_\textrm{min}$ for fitting constants to effective mass plateaus.}

\end{center}
\end{table}

The resulting mass differences $\Delta M(j^\mathcal{P}) a$ (in lattice
units), where \\ $j^\mathcal{P} \in \{ P_- \, , \, P_+ \, , \, D_\pm \,
, \, D_+ \, , \, F_\pm \, , \, S^\ast \}$, together with the pion masses
$m_\textrm{PS} a$ (in lattice units; cf.\ Table~\ref{TAB100} and
\cite{Baron:2009wt}) and the lattice spacings $a$ (in physical units;
cf.\ Table~\ref{TAB100}) serve as input for the extrapolation procedure
to physical $u/d$ quark masses described in the next section.

We checked the stability of our results by varying $T_\textrm{min}$ by
$\pm 1$ as well as by fitting superpositions of exponentials to the
elements of the correlation matrices (as done in \cite{Jansen:2008si})
instead of solving generalised eigenvalue problems. We found consistency
within statistical errors.



\section{Continuum limit and extrapolation to physical $u/d$ quark \\ masses}


\subsection{Numerical results}

\begin{figure}[p]
\begin{center}
\input{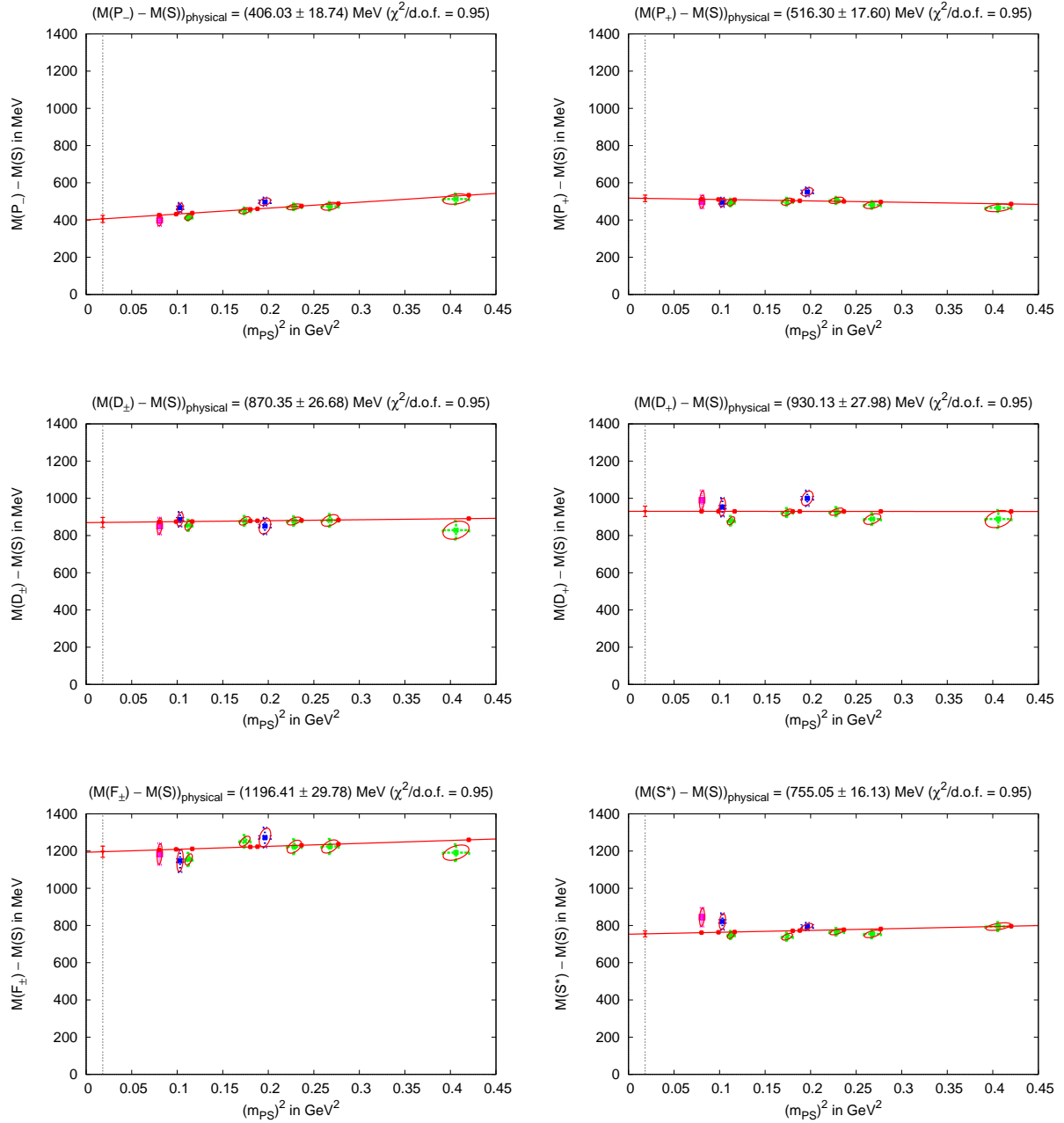}
\caption{\label{FIG001}static-light mass differences linearly extrapolated to the physical $u/d$ quark mass (unitary, i.e.\ $B$ mesons).}
\end{center}
\end{figure}

\begin{figure}[p]
\begin{center}
\input{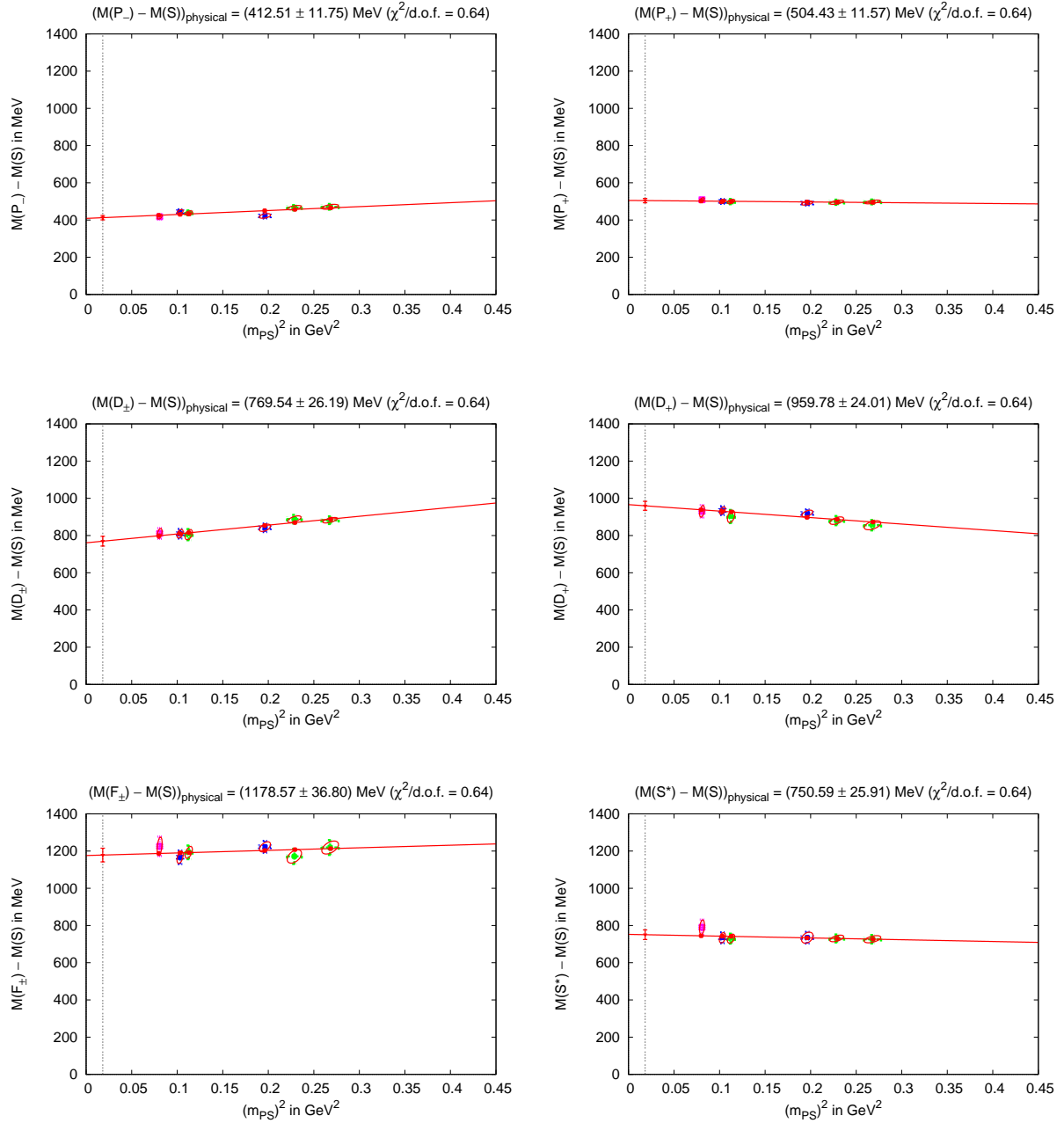}
\caption{\label{FIG002}static-light mass differences linearly extrapolated to the physical $u/d$ quark mass (partially quenched, i.e.\ $B_s$ mesons).}
\end{center}
\end{figure}

The mass differences $\Delta M(j^\mathcal{P})$ obtained for all our
ensembles are plotted against $(m_\textrm{PS})^2$ in Figure~\ref{FIG001}
(unitary, i.e.\ ``$B$ mesons'') and Figure~\ref{FIG002} (partially
quenched, i.e.\ ``$B_s$ mesons''). Note that, although we use
three different values of the lattice spacing, points corresponding to
the same mass difference fall on a single curve. This is reassuring,
since we use Wilson twisted mass lattice QCD at maximal
twist, where static-light mass differences are $\mathcal{O}(a)$
improved \cite{Jansen:2008si}. In Table~\ref{tab:eneB} and Table~\ref{tab:eneBs} we collect the values of the mass differences in $\textrm{MeV}$\footnote{The
scale has been set by the pion decay constant $f_\pi$ as explained in
detail in \cite{Baron:2009wt}.} for all simulation points for $B$ and $B_s$ mesons respectively.

\begin{table}[htb]
\begin{center}

\begin{tabular}{|c|c||c|c|c|c|c|c|}
\hline
 & & & & & & & \vspace{-0.40cm} \\
$\beta$ & $\mu_\mathrm{q}$ & $\Delta M (P_-)$  &  $\Delta M (P_+)$ & $\Delta M (D_\pm)$ & $\Delta M (D_+)$ & $\Delta M (F_\pm)$ & $\Delta M (S^*)$ \\
 & & & & & & & \vspace{-0.40cm} \\
\hline
 & & & & & & & \vspace{-0.40cm} \\
\hline
 & & & & & & & \vspace{-0.40cm} \\
$3.90$ &  $0.0040$ & $415(17)$ & $494(20)$ & $855(30)$ & $\phantom{0}879(25)$ & $1155(35)$ & $749(22)$ \\
       &  $0.0064$ & $449(17)$ & $499(20)$ & $879(26)$ & $\phantom{0}924(24)$ & $1253(33)$ & $740(21)$ \\
       &  $0.0085$ & $471(17)$ & $506(19)$ & $878(25)$ & $\phantom{0}928(24)$ & $1223(40)$ & $766(20)$ \\
       &  $0.0100$ & $474(22)$ & $481(21)$ & $881(34)$ & $\phantom{0}889(32)$ & $1225(40)$ & $755(23)$ \\
       &  $0.0150$ & $513(29)$ & $465(21)$ & $829(50)$ & $\phantom{0}889(48)$ & $1192(45)$ & $794(24)$ \\
 & & & & & & & \vspace{-0.40cm} \\
\hline
 & & & & & & & \vspace{-0.40cm} \\
$4.05$ &  $0.0030$ & $465(26)$ & $495(24)$ & $887(39)$ & $\phantom{0}952(49)$ & $1148(60)$ & $821(44)$ \\
       &  $0.0060$ & $498(22)$ & $551(23)$ & $851(44)$ & $          1000(41)$ & $1273(53)$ & $794(20)$ \\
 & & & & & & & \vspace{-0.40cm} \\
\hline
 & & & & & & & \vspace{-0.40cm} \\
$4.2$ &  $0.0020$ & $399(31)$ & $498(35)$ & $851(45)$ & $\phantom{0}990(53)$ & $1184(58)$ & $845(51)$\vspace{-0.40cm} \\
 & & & & & & & \\
\hline
\end{tabular}

\caption{\label{tab:eneB}static-light mass differences in $\textrm{MeV}$ (unitary, i.e.\ $B$ mesons) for all simulation points; details on the analysis procedure of the correlation functions are given in section~\ref{sec:energies}.}
\end{center}
\end{table}

\begin{table}[htb]
\begin{center}

\begin{tabular}{|c|c||c|c|c|c|c|c|}
\hline
 & & & & & & & \vspace{-0.40cm} \\
$\beta$ & $\mu_\mathrm{q}$ & $\Delta M (P_-)$  &  $\Delta M (P_+)$ & $\Delta M (D_\pm)$ & $\Delta M (D_+)$ & $\Delta M (F_\pm)$ & $\Delta M (S^*)$ \\
 & & & & & & & \vspace{-0.40cm} \\
\hline
 & & & & & & & \vspace{-0.40cm} \\
\hline
 & & & & & & & \vspace{-0.40cm} \\
$3.90$ &  $0.0040$ & $438(13)$ & $499(14)$ & $805(30)$ & $902(35)$ & $1193(37)$ & $729(26)$ \\
       &  $0.0085$ & $466(14)$ & $495(14)$ & $888(23)$ & $880(24)$ & $1171(41)$ & $730(21)$ \\
       &  $0.0100$ & $471(15)$ & $497(13)$ & $882(20)$ & $855(28)$ & $1219(40)$ & $726(22)$ \\
 & & & & & & & \vspace{-0.40cm} \\
\hline
 & & & & & & & \vspace{-0.40cm} \\
$4.05$ &  $0.0030$ & $444(13)$ & $500(13)$ & $810(26)$ & $934(24)$ & $1167(36)$ & $734(29)$ \\
       &  $0.0060$ & $422(14)$ & $491(13)$ & $842(23)$ & $918(22)$ & $1223(32)$ & $735(31)$ \\
 & & & & & & & \vspace{-0.40cm} \\
\hline
 & & & & & & & \vspace{-0.40cm} \\
$4.2$ &  $0.0020$ & $417(13)$ & $509(13)$ & $811(29)$ & $930(34)$ & $1226(52)$ & $790(41)$\vspace{-0.40cm} \\
 & & & & & & & \\
\hline
\end{tabular}

\caption{\label{tab:eneBs}static-light mass differences in $\textrm{MeV}$ (partially quenched, i.e.\ $B_s$ mesons)
for all simulation points; details on the analysis procedure of the correlation functions are given in section~\ref{sec:energies}.}
\end{center}
\end{table}

For the extrapolation to physical light quark masses, we could use an 
effective field theory approach (Chiral HQET for instance) as used to 
study the decay constants~\cite{Blossier:2009gd}  of the ground state.  This
approach   has not been developed to discuss mass differences between
excited states   and the ground state (e.g. $M(P_-) - M(S)$), so is not
appropriate here. Instead we use the simplest  assumption which is
supported by our results: a linear dependence.

 Because our ground state mass values enter into all of the mass
differences  we study, we simultaneously fit to all the meson mass
differences we have computed. We find that fits which are independent of
the lattice spacing and which are linear  in the light quark mass
(represented by the mass squared of the light-light  pseudoscalar meson)
are acceptable, i.e.\ yield $\chi^2 / \textrm{dof} \ltapprox 1$.

For the $B_s$ mesons, our results  depend on the strange quark mass 
we choose. We have taken these values from studies of strange-light 
mesons~\cite{Blossier:2007vv,Blossier:2009bx} as discussed above. The
possible systematic error arising from an incorrect  value for the
strange quark mass is very small: because the mass differences  we
measure turn out to be very weakly dependent on that mass. This will be
seen when we compare our results for the $B$ and $B_s$  mesons 
extrapolated to physical light quark masses.

The details of our fitting procedure are collected in appendix~\ref{APP001}.

As already mentioned both fits (one for $B$ mesons, the other for $B_s$ mesons) are of good quality in a sense that $\chi^2 / \textrm{dof} \ltapprox 1$. This shows that at the present level of statistical accuracy the continuum limit has already been reached at our largest value of the lattice spacing $a \approx 0.080 \, \textrm{fm}$. Moreover, these fits enable us to extrapolate to physical $u/d$ quark masses.

Extrapolations of static-light mass differences to physical $u/d$ quark
masses are listed in Table~\ref{TAB003} in $\textrm{MeV}$ both for $B$ mesons and for $B_s$
mesons. Note that both fits give $\chi^2 / \textrm{d.o.f.} \approx 1$,
i.e.\ are consistent with our assumption that static-light meson mass
differences as functions of $(m_\textrm{PS})^2$ can be parameterised by
straight lines.


\begin{table}[htb]
\begin{center}

\begin{tabular}{|c||c|c|c|c|c|c||c|}
\hline
 & & & & & & & \vspace{-0.40cm} \\
 & $P_-$ & $P_+$ & $D_\pm$ & $D_+$ & $F_\pm$ & $S^\ast$ & $\chi^2 / \textrm{d.o.f.}$ \\
 & & & & & & & \vspace{-0.40cm} \\
\hline
 & & & & & & & \vspace{-0.40cm} \\
\hline
 & & & & & & & \vspace{-0.40cm} \\
$B$ mesons   & $406(19)$ & $516(18)$ & $870(27)$ & $930(28)$ & $1196(30)$ & $755(16)$ & $0.95$ \\
 & & & & & & & \vspace{-0.40cm} \\
\hline
 & & & & & & & \vspace{-0.40cm} \\
$B_s$ mesons & $413(12)$ & $504(12)$ & $770(26)$ & $960(24)$ & $1179(37)$ & $751(26)$ & $0.64$\vspace{-0.40cm} \\
 & & & & & & & \\
\hline
\end{tabular}

\caption{\label{TAB003}$M(j^\mathcal{P}) - M(S)$ in $\textrm{MeV}$ extrapolated to physical light quark masses.}

\end{center}
\end{table}

To check the stability of these fits, we have varied $T_\textrm{min}$ by
$\pm 1$. Within statistical errors mass differences obtained with
$T_\textrm{min}-1$, with $T_\textrm{min}$ and with $T_\textrm{min}+1$
are in agreement.

The extrapolations are shown in Figure~\ref{FIG001} and
Figure~\ref{FIG002}. The red dots represent the maximum likelihood
estimates of $\bar{\mathbf{z}} = ((m_\textrm{PS})^2 \, , \, \Delta
M(j^\mathcal{P}))$ obtained during the fitting procedure. In addition to
$x$-$y$-error bars we also plot covariance ellipses, which reflect the
correlations between $(m_\textrm{PS})^2$ and $\Delta M(j^\mathcal{P})$
induced by the lattice spacing $a$, that is they are generated from the
inverses of the corresponding $2 \times 2$ submatrices of the covariance
matrix $C$.


\subsection{Contamination of static-light meson masses by multi particle states}

The radially and orbitally excited static-light mesons $P_-$, $P_+$,
$D_-$, $D_+$, $F_-$, $F_+$ and $S^\ast$ can decay into multi particle
states $S + n \times \pi$ with relative angular momentum such that
quantum numbers $j^\mathcal{P}$ are identical. In particular the $P_-$
static-light meson is not protected by angular momentum, i.e.\ it can
decay via an $S$ wave into $S + \pi$, whose wave function is not
suppressed at the origin. In the following we argue that the effect of
$S + \pi$ states on our $P_-$ mass is small compared to its statistical
error. To this end we resort to a model presented and to numerical
results obtained in \cite{McNeile:2000xx,McNeile:2002az,McNeile:2004rf}.

We consider the $P_-$ static-light meson at $\beta = 3.90$ and our
lightest $u/d$ quark mass at this $\beta$ value ($\mu_\mathrm{q} =
0.0040$). In that ensemble the masses of the $P_-$ state and of the $S +
\pi$ state are quite similar: $m(P_-) a \approx 0.57$ and $(m(S) +
m(\pi)) a \approx 0.53$ (we consider the case, where the pion has zero
momentum). Therefore, we expect  mixing of $P_-$ and $S + \pi$
with respect to the eigenstates of the Hamiltonian $H$, mixing which 
will be different in different spatial volumes. Consequently, we
do not focus on the eigenvalues of these states, but rather on
$m(P_-) = \langle P_- | H | P_- \rangle$ ($| P_- \rangle$ is a state
with $j^\mathcal{P} = (1/2)^+$ created by single particle operators,
e.g.\ operators of type $\bar{Q} u$ or $\bar{Q} d$, which we have used
in the construction of trial states). At very large temporal separation
the correlators we are studying will inevitably yield the eigenvalues of
the Hamiltonian. At intermediate temporal separations, however, one can
expect to read off $m(P_-)$ as we will explain in the following.

In \cite{McNeile:2004rf} the effective coupling strength of the decay
$P_- \to S + \pi$ has been estimated by a lattice computation: $\Gamma /
k \approx 0.46$. Moreover, some evidence has been obtained that this
quantity is fairly independent of the light quark mass. Using this
result one can determine the mixing element $x a$ of the energy matrix
via eqn.\ (5) in \cite{McNeile:2004rf} for our situation ($L/a = 24$, $m_\pi
a \approx 0.14$):
\begin{eqnarray}
x a \ \ = \ \ \bigg(\frac{2 \pi (\Gamma / k)}{ 3 (L/a)^3 (m_\pi a)}\bigg)^{1/2} \ \ \approx \ \ 0.023 .
\end{eqnarray}
 As detailed in \cite{McNeile:2000xx,McNeile:2002az,McNeile:2004rf} for
large temporal separations the $P_-$ correlator is of the form
\begin{eqnarray}
C_{P_-}(t/a) \ \ \propto \ \ e^{-(m_{P_-} a) (t/a)} \cosh((x a) (t/a)) ,
\end{eqnarray}
while the corresponding effective mass is
\begin{eqnarray}
\nonumber & & \hspace{-0.7cm} m_{\textrm{effective},P_-}(t/a) a \ \ = \ \ -\frac{d}{d(t/a)} \ln\Big(C_{P_-}(t/a)\Big) \ \ = \\
 & & = \ \ \frac{d}{d(t/a)} \bigg((m_{P_-} a) (t/a) - \ln\Big(\cosh((x a) (t/a))\Big)\bigg) \ \ = \ \ m_{P_-} a - \tanh((x a) (t/a)) x a .
\end{eqnarray}
 At $t/a = 12$ (the maximum temporal separation we have considered) the
estimated systematic error of $m_{P_-}$ coming from mixing with $S +
\pi$ is $\tanh((x a) (t/a)) x a \approx 0.0063$, i.e.\ roughly a $1 \%$
effect. This correction is significantly smaller than the statistical
error of $m_{\textrm{effective},P_-} a$ in that $t$ region.

For the other temporal separations and/or ensembles we obtain similar
estimates. We, therefore, expect that at the present level of
statistical accuracy the effect of multi particle states on our
static-light meson masses, in particular on $P_-$, is negligible.

Our conclusions are in agreement with those obtained in
\cite{Foley:2007ui}, where a study of the static-light meson spectrum
with similar techniques has been performed using two different lattice
volumes. No volume dependence of the eigenvectors of static-light meson
states has been observed, which is a sign that contributions of multi
particle states are strongly suppressed.



\section{Extrapolation to the physical $b$ quark mass}

%
%
%
%
%
%
%
%
%

To make contact with experimentally available results on the spectrum of
$B$ mesons, we need to correct for the non-infinite mass of the $b$
quark. In Heavy Quark Effective Theory, the leading correction will be
of order $1/m_H$, where $m_H$ is the heavy quark mass. It is possible,
in principle, to evaluate the coefficients of this correction  from
first principles on a lattice~\cite{Blossier:2009mg,Blossier:2010jk}.
 This we intend to explore in the  future, but here we use a more direct
method to establish the size of this small correction between static
quarks and $b$ quarks of realistic mass. These $1/m_H$ terms will break 
the degeneracy of mesonic states found in the static limit.

We evaluate for physical $b$ quarks by interpolating between static
heavy quarks and the charm quark, where experimental data is available.
As a measure of the heavy quark mass, we take the mass of the ground
state heavy-light meson ($D$ or $B$). This measure is equivalent to
another (such as using quark masses in some scheme) to the order $1/m_H$
we are using. One test of this interpolation can be made. The hyperfine
splitting between $D^*$ and $D$ of $141 \, \textrm{MeV}$ when
interpolated from the static limit (namely zero) gives for $B^*$ and $B$
a splitting reduced by \\ $m(D)/m(B) = 0.35$ to $49 \, \textrm{MeV}$
which agrees  with the observed splitting \cite{PDG} of $46 \,
\textrm{MeV}$ to within 6\%.

For the fine splitting, the kinetic term (rather than the
chromo-magnetic) is relevant and the experimental results for the 
spectrum are rather incomplete - indeed this current study is to establish 
the spectrum from a theoretical input.
 Lattice studies do confirm~\cite{Blossier:2009mg,Blossier:2010jk} that
a $1/m_H$ behavior is dominant down to masses near the  charm quark
mass.
 
\begin{figure}[p]
\begin{center}
\input{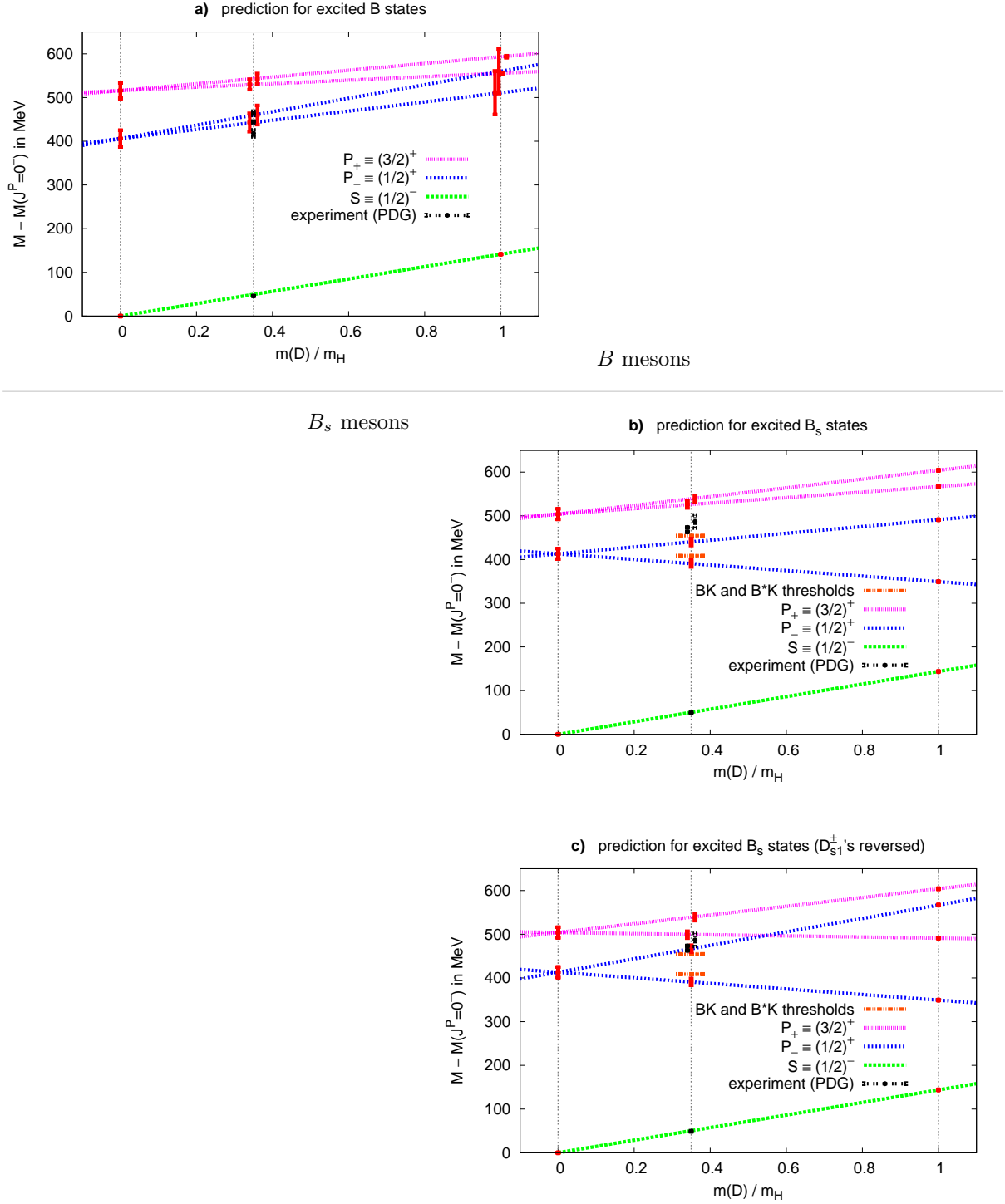}
\caption{\label{FIG003}Static-light mass differences linearly extrapolated to the physical $b$ quark mass.
\textbf{a)}~Unitary, i.e.\ $B$ mesons.
\textbf{b)}, \textbf{c)}~Partially quenched, i.e.\ $B_s$ mesons.
}
\end{center}
\end{figure}

We interpolate our lattice results for static-light mass differences of
$P$ and $S$ wave states to the physical $b$ quark mass at $m(D) / m(B) =
0.35$ linearly in $m(D)/m_H$, making  use of experimental data on $D$
and $D_s$ mesons as input \cite{PDG}. For details regarding this method
of extrapolation cf.\ \cite{Jansen:2008si}. Results are listed and
compared to experimental results in Table~\ref{TAB007}. The
corresponding extrapolations are shown in Figure~\ref{FIG003}.

For $D$ mesons the assignment of the two $J^\mathcal{P} = 1^+$ states to
$B_1^\ast$ and $B_1$ is easy, because their widths differ by
more than an order of magnitude (we associate the narrow state with
$B_1$ [one of the two degenerate $j^\mathcal{P}=(3/2)^+$ states in the
static limit, which can only decay to $S + \pi$ via a $D$ wave and is,
therefore, protected by angular momentum]; the wide state with $B_1^\ast$
[one of the two degenerate $j^\mathcal{P}=(1/2)^+$ states in the static
limit, which can readily decay to $S + \pi$ via an $S$ wave]). In
contrast to that the situation is less clear for $D_s$ mesons, where
both $J^\mathcal{P} = 1^+$ states have similar (narrow) widths.
Therefore, we show both possibilities in Table~\ref{TAB007} and in
Figure~\ref{FIG003}.

\begin{table}[htb]
\begin{center}
\begin{tabular}{|c|c|c||c|c|c|}
\hline
 & \multicolumn{2}{c||}{\vspace{-0.40cm}} & & \multicolumn{2}{c|}{} \\
 & \multicolumn{2}{c||}{$M-M(B)$ in MeV} & & \multicolumn{2}{c|}{$M-M(B_s)$ in MeV} \\
 & \multicolumn{2}{c||}{\vspace{-0.40cm}} & & \multicolumn{2}{c|}{} \\
\hline
 & & & & & \vspace{-0.40cm} \\
state & lattice & experiment & state & lattice & experiment \\
 & & & & & \vspace{-0.40cm} \\
\hline
 & & & & & \vspace{-0.40cm} \\
$B_0^\ast$ & $443(21)$ &          &   $B_{s0}^\ast$ & $391(8)$          & \\
$B_1^\ast$ & $460(22)$ &          &   $B_{s1}^\ast$ & $440(8) / 467(8)$ & \\
$B_1$      & $530(12)$ & $444(2)$ &   $B_{s1}$      & $526(8) / 499(8)$ & $463(1)$ \\
$B_2^\ast$ & $543(12)$ & $464(5)$ &   $B_{s2}^\ast$ & $539(8)$          & $473(1)$ \\
 & & & & & \vspace{-0.40cm} \\
\hline
 & & & & & \vspace{-0.40cm} \\
$B_J^\ast$ &           & $418(8)$ &   $B_{sJ}^\ast$ &           & $487(15)$\vspace{-0.40cm} \\
 & & & & & \\
\hline
\end{tabular}

 \caption{\label{TAB007}lattice and experimental results for $P$ wave
$B$ and $B_s$ states ($B_J^\ast$ and $B_{sJ}^\ast$ denote rather vague
experimental signals, which can be interpreted as stemming from several
broad and narrow resonances possibly including the $j=1/2$ $P$ wave
states $B_0^\ast$, $B_1^\ast$, $B_{s0}^\ast$ and $B_{s1}^\ast$; the two
lattice values listed for $B_{s1}^\ast$ and $B_{s1}$ correspond to the
two possibilities of assigning experimental $J^\mathcal{P} = 1^+$ $D$
results [cf.\ text for more details]).}
 \end{center}
\end{table}

Compared to our previous study \cite{Jansen:2008ht,Jansen:2008si} at a
single lattice spacing, the above results are similar for the $B$
(unitary) case. For $B_s$ mesons we now employ a partially quenched $s$
quark which  allows a more realistic treatment of the light quark sea.
So our new results  supersede those obtained previously for $B_s$.
Indeed we find a significant dependence on the sea quark mass (cf.\
Figure~\ref{FIG002}), which is now the physical $u/d$ quark mass, while
it previously corresponded to  the significantly heavier $s$ quark mass.

In our lattice study we have extracted the continuum limit and have 
extrapolated to physical light quarks using a linear dependence. We have
then  interpolated to the physical $b$ quark assuming that a $1/m_H$
behavior is valid  down to the charm quark mass. These assumptions
induce systematic errors and,  in principle, they can be quantified by
further lattice studies. 

The assumption of a linear extrapolation to physical light quarks is 
sensitive to possible admixtures of two body states which become  more
important at lighter quark masses as thresholds for decay open. We have 
explored this possibility and found no evidence of such effects, so it
is  difficult to estimate the magnitude of a possible systematic error
from this. If there was a significant difference between the light quark 
behavior for the ground state and an excited state, this would introduce 
an error on our extrapolation to the physical value which could be as 
large as $10 \, \textrm{MeV}$.

The test  of the $1/m_H$ assumption for the chromo-magnetic term,
discussed above,  was found to be valid within 6\%. This suggests that
an estimate of the systematic errors for the $B$ and $B_s$ meson mass
splittings coming from  $1/m$ effects should also be at least of order
6\%. Since the $1/m$ correction to the $P$ wave states is of order $100
\, \textrm{MeV}$, this implies a systematic error of order $6 \,
\textrm{MeV}$.

One further possible source of systematic error is from our neglect of
the  strange contribution to the sea. This will be addressed in a future
study making use of the $N_f=2+1+1$ sea which includes 
dynamical $s$ quarks from ETMC \cite{Baron:2008xa,Baron:2009zq}.

Overall, it seems prudent to assign systematic errors on our mass 
differences (for $P_-$ and  $P_+$ relative to $S$) of order $20 \, \textrm{MeV}$ from these  effects, even though we have little evidence for such effects.

The experimental determination of the spectrum of excited $B$ and $B_s$
mesons is quite limited \cite{PDG}. Assuming that the relatively narrow
states seen correspond to our $P_+$ state (since a $J^\mathcal{P} = 2^+$
state must have that assignment), the mass difference we see of over
$500 \, \textrm{MeV}$ does not agree closely with the experimental
results of around $450 \, \textrm{MeV}$. We do get a mass difference of
around $450 \, \textrm{MeV}$ from our $P_-$ states, although such states
cannot have $J^\mathcal{P} = 2^+$.

In view of this discrepancy with experimental results, it is also
interesting to compare with independent existing lattice computations,
in particular with the rather recent study reported in
\cite{Burch:2008qx}. There the light quark extrapolation is only
performed in the valence quark mass (from which static-light mass
differences essentially seem to be independent, as can be seen by
comparing our $B$ and $B_s$ results and also from corresponding plots
and numbers presented in \cite{Burch:2008qx}), while the sea quark mass
is kept fixed. More generally, a comparison of the dependence of
static-light mass differences on the sea quark mass, which we have
computed down to $m_\textrm{PS} \approx 280 \, \textrm{MeV}$, with
existing lattice studies is not possible: there the number of
investigated sea quark masses is rather small and they are quite heavy,
around the mass of the $s$ quark. What one can do, however, is to
compare meson mass differences for a given value of the sea quark mass.
 Before comparing results (in physical units) with those quoted in
\cite{Burch:2008qx} it should be noted that in \cite{Burch:2008qx} the
scale is set by identifying $r_0$ with $0.49 \, \textrm{fm}$, while our
result for this quantity is $r_0 = 0.42 \, \textrm{fm}$
\cite{Baron:2009wt}. Therefore, to perform a meaningful comparison, one
should express all quantities in units of $r_0$ or equivalently scale
all masses in physical units listed in \cite{Burch:2008qx} by a factor
of around $0.49 / 0.42 \approx 1.14$. For the lightest sea quark mass
considered in \cite{Burch:2008qx} corresponding to $m_\textrm{PS}
\approx 461 \, \textrm{MeV}$ it is most appropriate to compare with our
results at $\beta = 3.90$, $\mu_\mathrm{q} = 0.0100$ ($m_\textrm{PS}
\approx 517 \, \textrm{MeV}$). For the $P$ wave mass differences one
finds
\begin{eqnarray}
 & & \hspace{-0.7cm} \frac{(m(P_-) - m(S))_\textrm{ETMC}}{(m(P_-) - m(S))_\textrm{\cite{Burch:2008qx}}} \ \ \approx \ \ \frac{474(29) \, \textrm{MeV}}{454(19)(9) \, \textrm{MeV}} \ \ \approx \ \ 1.04(11) \\
 & & \hspace{-0.7cm} \frac{(m(P_+) - m(S))_\textrm{ETMC}}{(m(P_+) - m(S))_\textrm{\cite{Burch:2008qx}}} \ \ \approx \ \ \frac{481(27) \, \textrm{MeV}}{446(17)(9) \, \textrm{MeV}} \ \ \approx \ \ 1.08(11) ,
\end{eqnarray}
ratios, which are within statistical errors fully consistent with the
expected factor $1.14$.

It is interesting to note that the ratios between our lattice results and the experimental values (see Table~\ref{TAB007}) are on the same ballpark of the ratio between two values of $r_0$ used above, i.e.\ $\approx 1.14$. While there is no reason to doubt the precise determination of the lattice spacing performed in \cite{Baron:2009wt}, it would be interesting, although beyond the scope of this paper, to investigate, whether simulations at lighter quark masses and/or with $N_f = 2+1+1$ dynamical flavours will improve the agreement with experimental results.

One interesting issue is whether the $B_s$ states are stable to the
strong  decay to $B K$. This decay has a threshold at $408 \,
\textrm{MeV}$ above the ground state $B_s$ meson. Our $P_-$ states (the
upper two in Table~\ref{TAB007}) do indeed have  masses which are close
to (or below) this threshold. That would imply that  these two states
($B^*_{s0}$ and $B^*_{s1}$) should have a very small decay width. This
is consistent with the experimental observation that only two candidate
$P$ wave $B^*$ states have been seen so far: corresponding to the heavier
$P_+$ states. All the other states $B_s$ we study, including the $S^*$,
lie higher than this $B K$ threshold and so would have a strong decay
open.

Moreover, our findings clearly indicate that there is no inversion of
level ordering for $P$ wave states, neither for $B$ mesons nor for $B_s$
mesons. $B_0^\ast$ and $B_1^\ast$ ($B_{s0}^\ast$ and $B_{s1}^\ast$) are
considerably lighter than $B_1$ and $B_2^\ast$ ($B_{s1}$ and
$B_{s2}^\ast$) as can be read off from Table~\ref{TAB007} and
Figure~\ref{FIG003}. This is in contrast to predictions obtained from
certain phenomenological models
\cite{Schnitzer:1978gq,Schnitzer:1989xr,Ebert:1997nk,Isgur:1998kr,Ebert:2009ua}
and, therefore, might provide valuable input for future model building.



\section{Conclusions}

We have determined the continuum limit for static-light mesons on a
lattice using $N_f=2$ flavours of light quarks. The removal of
$\mathcal{O}(a)$ effects by using maximally-twisted mass fermions for
meson mass differences in the static limit is confirmed.

We have investigated the light sea quark mass dependence of $B$ and
$B_s$ mesons down to \\ $m_\textrm{PS} \approx 280 \, \textrm{MeV}$,
which is significantly lighter than what has been achieved in previous
studies of static-light mesons. We find that our results are compatible
with a linear extrapolation in the light quark mass to its physical
value. We see no sign of any mixing with two body effects and this is
consistent with our estimate that such effects should be too small to
see on our lattices.

We have determined masses for a wide variety of excited states in the
continuum limit and this will be a valuable resource for model builders.

We have employed the assumption of a $1/m_H$ dependence on the heavy
quark mass together with experimental results for charm-light mesons to
allow us to estimate the spectrum that one would obtain for physical $b$
quarks.

Our results imply that there will be a $J^\mathcal{P} = 0^+$ and
$J^\mathcal{P} = 1^+$ $B_s$ meson which has a narrow width since its
strong decay to $B K$ is suppressed (or zero) due to phase space
effects.

Future directions include (i)~determination of $f_B$ and $f_{B_s}$ (for
a preliminary result cf.\ \cite{Blossier:2009gd}); (ii)~a similar
investigation regarding static-light baryons; (iii)~extending these
computations to $N_f = 2+1+1$ flavour ETMC gauge configurations
\cite{Baron:2008xa,Baron:2009zq}.


\appendix


\section{\label{APP001}Details of the fitting procedure}

Data points $(m_\textrm{PS})^2$ and $\Delta M(j^\mathcal{P})$,
$j^\mathcal{P} \in \{ P_- \, , \, P_+ \, , \, D_\pm \, , \, D_+ \, , \,
F_\pm \, , \, S^\ast \}$ corresponding to the same $\beta$ are
correlated via the lattice spacing $a$. We take that into account via a
covariance matrix, which we estimate by resampling $m_\textrm{PS} a$,
$\Delta M(j^\mathcal{P}) a$ and $a$ ($100,000$ samples). Consequently,
we do not fit straight lines to the Data points $((m_\textrm{PS})^2 \, ,
\, \Delta M(j^\mathcal{P}))$ individually for every static-light state
$j^\mathcal{P}$, but perform a single correlated fit of six straight
lines to the six mass differences of interest. During the fitting we
take statistical errors both along the horizontal axis (errors in
$(m_\textrm{PS})^2$) and along the vertical axis (errors in $\Delta
M(j^\mathcal{P})$) into account.

The method of performing the two-dimensional fits is based on what has
been used in \cite{Farchioni:2005bh}.

To be able to express the corresponding equations in a compact way, we
introduce the following notation:
\begin{itemize}
\item $\mathbf{z} = (\mathbf{x} \, , \, \mathbf{y}(1) \, , \, \mathbf{y}(2) \, , \, \ldots)$.

\item $\mathbf{x} = (((m_\textrm{PS})^2)^{(1)} \, , \, ((m_\textrm{PS})^2)^{(2)} \, , \, \ldots)$ (the upper index $^{(\ldots)}$ refers to both the lattice spacing and to the light quark mass).

\item $\mathbf{y}(j) = ((\Delta M)^{(1)}(j) \, , \, (\Delta M)^{(2)}(j) \, , \, \ldots)$ (the upper index $^{(\ldots)}$ refers to both the lattice spacing and to the light quark mass, the index $(j)$ refers to $j^\mathcal{P}$).

\item $C$ denotes the estimated covariance matrix for $\mathbf{z}$ (a $56 \times 56$ matrix for $B$ mesons, a $42 \times 42$ matrix for $B_s$ mesons).

\item The linear fits $y(j) = a(j) x + b(j)$ are parameterised by $a(j)$ and $b(j)$ (the quantities, which will finally allow the extrapolation to physical $u/d$ quark masses).
\end{itemize}

The basic idea of the method is a maximum likelihood determination of the ``true values'' \\ $\bar{\mathbf{z}} = (\bar{\mathbf{x}} \, , \, \bar{\mathbf{y}}(1) \, , \, \bar{\mathbf{y}}(2) \, , \, \ldots)$. This amounts to minimizing
\begin{eqnarray}
\frac{1}{2} \Big(\mathbf{z} - \bar{\mathbf{z}}\Big)^T C^{-1} \Big(\mathbf{z} - \bar{\mathbf{z}}\Big) - \sum_{j,n} \lambda_n(j) \Big(a(j) \bar{x}_n + b(j) - \bar{y}_n(j)\Big)
\end{eqnarray}
with respect to $\bar{\mathbf{z}}$, $a(j)$, $b(j)$ and $\vec{\lambda}(j)$ under the constraints $\bar{y}_n(j) = a(j) \bar{x}_n + b(j)$.
%
%
%
%
%
%
%
%
For $\mathbf{z}$ we use the same resampling procedure as for estimating the covariance matrix (this is necessary, because $z_A \equiv \langle ((m_\textrm{PS})^2)^{(n)} \rangle \neq \langle (m_\textrm{PS})^{(n)} a\rangle^2 / \langle a \rangle^2$ and $z_A \equiv \langle (\Delta M)^{(n)}(j) \rangle \neq \langle (\Delta M)^{(n)}(j) a \rangle / \langle a \rangle$).

The constraint minimization is equivalent to solving a system of non-linear equations, which we do by means of the scaled-hybrid algorithm of the GSL library \cite{GSL}. It needs initial parameters, which should preferably be close to the global extremum. Such initial parameters can be obtained by individual standard one-dimensional straight line fits:
\begin{itemize}
\item $\lambda_n(j) = 0$,

\item $a(j)$ and $b(j)$ minimizing
\begin{eqnarray}
\sum_n \frac{\Big(a(j) x_n + b(j) - y_n(j)\Big)^2}{C_{y_n(j),y_n(j)}} ,
\end{eqnarray}

\item $\bar{\mathbf{x}} = \mathbf{x}$ and $\bar{\mathbf{y}}(j) = \mathbf{y}(j)$.
\end{itemize}

To judge the quality of the resulting fit, we define a ``reduced $\chi^2$'' via
\begin{eqnarray}
\frac{\chi^2}{\textrm{d.o.f.}} \ \ = \ \ \frac{\Big(\mathbf{z} - \bar{\mathbf{z}}\Big)^T C^{-1} \Big(\mathbf{z} - \bar{\mathbf{z}}\Big)}{\textrm{d.o.f.}} ,
\end{eqnarray}
 where $\textrm{d.o.f.}$ is the number of entries of $\bar{\mathbf{z}}$
minus the number of $a(j)$ and $b(j)$, i.e.\ $\textrm{d.o.f.} = 44$ for
$B$ mesons and $\textrm{d.o.f.} = 32$ for $B_s$ mesons respectively.

The resulting straight lines allow an extrapolation to physical $u/d$
quark masses (corresponding to $m_\textrm{PS} = 135 \, \textrm{MeV}$).
The corresponding statistical errors are obtained by repeating this
fitting and extrapolation procedure $100$ times with randomly sampled
sets $z_A$ (we randomly sample the input data and compute $z_A \equiv
((m_\textrm{PS})^{(n)} a)^2 / a^2$ and $z_A \equiv ((\Delta M)^{(n)}(j)
a) / a$) and taking the variance.


\section*{Acknowledgments}

We thank R\'emi Baron for running contractions at $\beta = 4.20$. We acknowledge useful discussions with Vladimir Galkin, Karl Jansen, Marcus Petschlies and Carsten Urbach.

A.S.\ acknowledges financial support from Spanish Consolider-Ingenio 2010 Programme CPAN (CSD 2007-00042) and from Comunidad Aut\'onoma de Madrid, CAM under grant HEPHACOS P-ESP-00346. This work has been supported in part by the DFG Sonderforschungsbereich/Transregio SFB/TR9-03.

This work was performed using HPC resources from GENCI/IDRIS Grant 2009-052271. We thank CCIN2P3 in Lyon and the J\"ulich Supercomputing Center (JSC) for having allocated to us computer time, which was used in this work. We acknowledge computing resources provided by the NW Grid at Liverpool.



\end{document}